\documentstyle[aps,preprint,tighten,psfig]{revtex}

\newcommand{\lsim}{\mathrel{\mathop{\kern 0pt \rlap
  {\raise.2ex\hbox{$<$}}}
  \lower.9ex\hbox{\kern-.190em $\sim$}}}
\newcommand{\gsim}{\mathrel{\mathop{\kern 0pt \rlap
  {\raise.2ex\hbox{$>$}}}
  \lower.9ex\hbox{\kern-.190em $\sim$}}}
\newcommand{\beq}    {\begin{equation}}
\newcommand{\eeq}    {\end{equation}}
\newcommand{\beqarr} {\begin{eqnarray}}
\newcommand{\eeqarr} {\end{eqnarray}}
\newcommand{\barr}   {\begin{array}}
\newcommand{\earr}   {\end{array}}
\newcommand{\no}     {\nonumber}

\newcommand{\mb}[1]  {\mbox{#1}}

\begin{document}


\title{Probing the supersymmetric parameter space by WIMP direct
  detection}

\author{\bf 
A. Bottino$^{\mbox{a}}$
\footnote{E--mail: bottino@to.infn.it, donato@lapp.in2p3.fr, 
fornengo@to.infn.it, \\
\phantom{E--mail:~~~} scopel@to.infn.it},
F. Donato$^{\mbox{b}}$\footnote[4]{INFN Post--doctoral Fellow}, 
N. Fornengo$^{\mbox{a}}$,
S. Scopel$^{\mbox{a}}$
\vspace{6mm}
}

\address{
\begin{tabular}{c}
$^{\mbox{a}}$
Dipartimento di Fisica Teorica, Universit\`a di Torino \\
and INFN, Sez. di Torino, Via P. Giuria 1, I--10125 Torino, Italy\\
$^{\mbox{b}}$
Laboratoire de Physique  Th\'eorique LAPTH, B.P. 110, F--74941\\
Annecy--le--Vieux Cedex, France \\
and INFN, Sede di Presidenza, 00186 Roma, Italy
\end{tabular}
}

\maketitle

\begin{abstract}
  We discuss to which extent the present experiments of direct search for
  WIMPs, when interpreted in terms of relic neutralinos, probe interesting
  regions of the supersymmetric parameter space, which are also being
  progressively explored at accelerators.  Our analysis is performed in a
  number of different supersymmetric schemes. We derive the relevant neutralino
  cosmological properties, locally and on the average in the universe.  We
  prove that part of the susy configurations probed by current WIMP experiments
  entail relic neutralinos of cosmological interest.  The main astrophysical
  and particle physics uncertainties, relevant for a proper comparison between
  theory and experimental data, are stressed and taken into account.
\end{abstract}  

\vspace{1cm}


\section{Introduction}  

As was first noticed in Ref. \cite{bdmsbi}, in the last few years the
experiments of direct search for Weakly Interacting Massive Particles
(WIMP) \cite{morales} have already reached a sensitivity which allows
the exploration of regions of the supersymmetric parameter space,
which are also progressively investigated at accelerators.  This
property is manifest, when the experimental results are interpreted in
terms of relic neutralinos \cite{bdmsbi}.

The probing of the susy parameter space by WIMP direct searches is
even more sizeable at present, with the sensitivities of experiments
\cite{damalast,cdms}; a comparative discussion of the experimental
features and implications of the DAMA \cite{damalast} and CDMS
\cite{cdms} experiments may be found in Ref. \cite{damapr}.  Detailed
studies of the possible interpretation of the annual--modulation
effect \cite{damalast} in terms of relic neutralinos have been
reported in \cite{noi,comp,noi5,noiult}. Comparisons of the
experimental data of Ref. \cite{damalast} with susy calculations have
also been performed in Refs. \cite{an,kkk,efo,acc,gabr,el2}.

In the present paper we intend to clarify the actual capability of
WIMP direct searches by exploring in a sistematic way different
realizations of minimal supersymmetric models and showing their
intrinsic differences in the prediction of neutralino rates and relic
abundance.  Specifically, we will consider two different
implementations of a supergravity scheme with parameters defined at
the grand unification scale and an effective supersymmetric model
defined at the electroweak scale.  Our analyses are performed in the
light of the following relevant points: i) current uncertainties in
astrophysical properties, ii) uncertainties in hadronic quantities,
iii) new bounds from LEP searches for Higgs and supersymmetric
particles, iv) updated determinations of cosmological parameters.

Let us start by recalling that  the
 determination of the sensitivity range of an experiment of WIMP 
direct search {\it in terms of the WIMP mass and of the WIMP--nucleon 
cross section} rests on a number of crucial assumptions, since it 
depends both on the distribution function of the WIMPs in the halo and 
on the nature of the relic particle. 

       A WIMP direct experiment provides a measurement (or an 
upper bound)  of the differential event rate 

\begin{equation}
\frac {dR}{dE_R}=N_{T}\frac{\rho_{W}}{m_{W}}
                    \int \,d \vec{v}\,f(\vec v)\,v
                    \frac{d\sigma}{dE_{R}}(v,E_{R}) 
\label{eq:diffrate0}
\end{equation}

\noindent
where $N_T$ is the number of the target nuclei per unit of mass, $m_W$
is the WIMP mass, $\rho_W$ is the local WIMP matter density, $\vec v$
and $f(\vec v)$ denote the WIMP velocity and velocity distribution
function in the Earth frame ($v = |\vec v|$) and $d\sigma/dE_R$ is the
WIMP--nucleus differential cross section.  The nuclear recoil energy
is given by $E_R={{m_{\rm red}^2}}v^2(1-\cos \theta^*)/{m_N}$, where
$\theta^*$ is the scattering angle in the WIMP--nucleus
center--of--mass frame, $m_N$ is the nuclear mass and $m_{\rm red}$ is
the WIMP--nucleus reduced mass.  Eq.(\ref{eq:diffrate0}) refers to the
case of a monoatomic detector, like the Ge detectors. Its
generalization to more general situations, like for instance the case
of NaI, is straightforward. In what follows $\rho_W$ will be
factorized in terms of the local value for the total non--baryonic
dark matter density $\rho_l$ and of the fractional amount of density,
$\xi$, contributed by the candidate WIMP, {\it i.e.} $\rho_W = \xi \cdot
\rho_l$. For $\rho_l$ we use the range 0.2 GeV cm$^{-3} \leq \rho_l
\leq$ 0.7 GeV cm$^{-3}$, where the upper side of the range takes into
account the possibility that the matter density distribution is not
spherical, but is described by an oblate spheroidal distribution
\cite{bt,t}.

The WIMP--nucleus differential cross section may conveniently be split
into a coherent part and a spin--dependent one

\begin{equation}
\frac {d\sigma}{dE_R} = \left(\frac{d\sigma}{d E_R}\right)_C+
                        \left(\frac{d\sigma}{d E_R}\right)_{SD}, 
\label{eq:diffrate_sum}
\end{equation}

\noindent
 whose generic features are discussed
in the seminal paper of Ref. \cite{gw}.
 To compare theoretical expectations with experimental data, and
experimental data of different detectors among themselves, it is
useful to convert the WIMP--nucleus cross--section into a
WIMP--nucleon cross section. This procedure is feasible independently of
the nuclear model and of the specific nature of the WIMP  only
under the hypothesis that the coherent cross--section is dominant and 
 the WIMP couples equally  to protons and neutrons 
(at least approximately) \cite{bdmsbi}. 
  Under this assumption, the   WIMP--nucleus cross section
may be expressed in terms of a WIMP--nucleon scalar cross section
 $\sigma^{\rm (nucleon)}_{\rm scalar}$  as 
   
\begin{equation}
\frac {d\sigma}{dE_R} \simeq \left ( \frac{d\sigma}{dE_R} \right )_C
              \simeq \frac{F^2(q)}{E^{max}_R}
              \left(\frac{1+m_W/m_p}{1+m_W/m_N}\right )^2 A^2
              \sigma^{\rm (nucleon)}_{\rm scalar}, 
\label{eq:diffrate_approx}
\end{equation}

\noindent where $m_p$ and $m_N$ are the proton and nucleus mass,
$A$ is the nuclear mass number, $E_R^{max}$ is the maximal recoil
energy and $F(q)$ is the nuclear form factor for coherent
interactions.
This form factor is  usually parametrized in the Helm form \cite{helm};
however, precise evaluations of the event rates may require
specific nuclear calculations for each target nucleus.
In the rest of this paper we assume that the WIMP interaction with the
nuclei of the detector is dominated by coherent effects, so that a 
WIMP--nucleon scalar cross section may be derived from the WIMP--nucleus
cross section by use of Eq. (\ref{eq:diffrate_approx}).

Now, coming back  to the general expression in Eq. (\ref{eq:diffrate0}), 
we stress  that  
extracting an information about the WIMP--nucleus cross sections from 
the experimental data requires the use of a specific expression for 
the velocity  distribution function $f(\vec{v})$ (notice that in writing 
  Eq. (\ref{eq:diffrate_approx}) we have already made the
 assumption that the WIMP 
phase--space distribution function may be factorized as 
$\rho \cdot f(\vec{v})$, and this is certainly not the most general case 
\cite{bt}).   
  The usual choice for $f(\vec{v})$ is the isotropic  Maxwell--Boltzmann 
distribution in the galactic rest frame, as derived from the 
isothermal-sphere model. 

However, recent investigations have shown that deviations from this
standard scheme, either due to a bulk rotation of the dark halo
\cite{kk,dfs} or to an asymmetry in the WIMP velocity distribution
\cite{vu,ecz,amg}, influence the determination of the WIMP--nucleus
cross sections from the experimental data in a sizeable way.  In Ref.
\cite{ecz} also triaxial matter distributions are considered; in the
present paper deviation from sphericity in the WIMP matter
distributions are taken into account only through the physical range
allowed for the value of $\rho_l$ (see our previous comment on
$\rho_W$ after Eq. (\ref{eq:diffrate0})).  In a typical plot,
where the WIMP-nucleus cross section is given in terms of the WIMP
mass, the effect introduced by the mentioned deviations from the
Maxwell--Boltzmann distribution is generically to elongate the contours towards
larger values of $m_W$. This is for instance the case for the the
annual--modulation region of the DAMA Collaboration \cite{damalast}.
In Fig. 3 of Ref. \cite{noi5} it is shown that, by implementing the
dark halo with a bulk rotation according to the treatment in Ref.
\cite{dfs}, the annual--modulation region moves towards larger values
of the WIMP mass, with an elongation which brings the right--hand
extreme from the value of $\, \sim$ 150 GeV to $\, \sim$ 200 GeV.  A
similar effect is obtained by introducing an asymmetry in the WIMP
velocity distribution $f(\vec{v})$: Fig. 4 of Ref. \cite{amg}
illustrates this point. Notice that this asymmetry effect also pushes
somewhat downwards the annual--modulation region.  We emphasize that
all these effects are extremely important, when experimental results
of WIMP direct detection are being compared with theoretical models
for specific candidates.  This point has been overlooked in most
analyses in terms of relic neutralinos \cite{caustics}.

In the present paper we focus  our analysis to the WIMP mass
range which, in the light of the present experimental data
\cite{damalast,cdms} and of the previous considerations on the
astrophysical uncertainties, appears
particularly appealing:  

\begin{equation}
40 \; {\rm GeV} \leq  m_W \leq 200 \;  {\rm GeV}. 
\label{eq:mass}
\end{equation}

Let us notice that the mass range of Eq. (\ref{eq:mass}) is 
quite appropriate for neutralinos. Actually, the lower extreme 
is indicative of the LEP lower bound on the neutralino mass 
$m_{\chi}$ (in the calculations performed in the present work the
actual lower bound for $m_{\chi}$, dependent on the other susy 
parameters, is employed, according to the
constraints given in \cite{LEPb}). 
As for the upper 
extreme, we notice that, though a generic  range for $m_{\chi}$
 might extend up to about 1 TeV, requirements of no excessive 
fine--tuning  \cite{bere1}
would actually  favour an upper bound of order 200 GeV, 
in accordance with Eq. (\ref{eq:mass}).  

In what follows we will discuss the discovery potential of WIMP direct
searches for WIMPs in the mass range of Eq.(\ref{eq:mass}). Particular
attention will be paid to capabilities of the present experiments;
their sensitivity range, in case of WIMPs whose coherent interactions
with ordinary matter are dominant over the the spin--dependent ones,
may be stated, in terms of the quantity $\xi \sigma^{\rm
  (nucleon)}_{\rm scalar}$, as \cite{damalast,cdms}

\begin{equation}
4 \cdot 10^{-10} \; {\rm nbarn} \leq \
\xi \sigma^{\rm (nucleon)}_{\rm scalar} \leq 
 2 \cdot 10^{-8} \; {\rm nbarn}.
\label{eq:section}
\end{equation}

We will hereafter refer to region $R$ as the one in the space 
$m_W - \xi \sigma^{\rm (nucleon)}_{\rm scalar}$ which is  defined by 
Eqs. (\ref{eq:mass}-\ref{eq:section}). The region $R$ represents the
sensitivity region already under exploration with present
detectors.

Our analysis, based on an interpretation of experimental data in
terms of relic neutralinos,  will show by how much the
WIMP direct searches probe the supersymmetric parameter space. 
We remark  that, in the case of neutralinos, the assumption about the 
dominance  of the coherent cross section over the spin--dependent one 
is, in general, largely satisfied, except for values of 
 $\sigma^{\rm (nucleon)}_{\rm scalar}$ which are 
far below  the present   experimental reach \cite{bdmsbi}.

The present analysis will be performed in the framework of various
schemes, from those based on universal or non-universal supergravity,
with susy parameters defined at the grand unification scale, to an
effective supersymmetric model defined at the Electro--Weak (EW)
scale. This is discussed in Sect. II, where we also specify the values
employed here for the Higgs--quark--quark and the
neutralino--quark--squark couplings.  These quantities are subject to
sizeable uncertainties, as was stressed in Ref. \cite{noi6}, which
triggered a reconsideration of this important point in a number of
subsequent papers \cite{efo,acc,cn}.

The most important properties to be established for the relic
neutralinos, which are entailed in the exploration by WIMP direct
searches, concern their cosmological properties. Here, we perform a
general analysis which is not limited to a restricted range of the
cosmological matter abundance, $\Omega_m h^2$ ($\Omega_m$ is the
matter cosmological density divided by the critical density and $h$ is
the Hubble constant in units of 100 km s$^{-1}$ Mpc$^{-1}$) . Instead,
we derive the average and local cosmological properties of the susy
configurations from experimental determinations of $\sigma^{\rm
  (nucleon)}_{\rm scalar}$, without any {\it a priori} requested range
on $\Omega_m h^2$. On the basis of the results of our evaluations in
the various supersymmetry models, we discuss when the relic neutralino
does or does not saturate the expected amount of the local and of the
average amount of {\it total} dark matter.  Our results and
conclusions are presented in Sects. III and IV, respectively.
 
\section{Supersymmetric Models}

The calculations presented in this paper are based on the Minimal
Supersymmetric extension of the Standard Model (MSSM), in a variety of
different schemes.  The essential elements of the MSSM are described
by a Yang--Mills Lagrangian, the superpotential, which contains all
the Yukawa interactions between the standard and supersymmetric
fields, and by the soft--breaking Lagrangian, which models the
breaking of supersymmetry.  To fix the notations, we write down
explicitly the soft supersymmetry breaking terms
\begin{eqnarray}
&-{\cal L}_{soft}& =
\displaystyle \sum_i m_i^2 |\phi_i|^2
\no \\
&+&  \left\{\left[
A^{l}_{ab} h_{ab}^{l} \tilde L_a H_1 \tilde R_b +
A^{d}_{ab} h_{ab}^{d} \tilde Q_a H_1 \tilde D_b +
A^{u}_{ab} h_{ab}^{u} \tilde Q_a H_2 \tilde U_b +\mb{h.c.} \right] -
B \mu H_1 H_2 + \mb{h.c.}      \right\}
\no \\
&+& \displaystyle \sum_i M_i
(\lambda_i \lambda_i + \bar\lambda_i \bar\lambda_i)
\label{eq:soft}
\end{eqnarray}
\noindent
where the $\phi_i$ are  the scalar fields, the $\lambda_i$ are the
gaugino fields, $H_1$ and $H_2$ are the two Higgs fields,
$\tilde Q$ and $\tilde L$
are the doublet squark and slepton fields, respectively,
and $\tilde U$, $\tilde D$ and
$\tilde R$ denote the $SU(2)$--singlet fields for the up--squarks,
down--squarks and sleptons. In Eq.(\ref{eq:soft}), $m_i$ and $M_i$ are the mass
parameters of the scalar and gaugino fields, respectively, and $A$ and
$B$ denote trilinear and bilinear supersymmetry breaking parameters,
respectively. The
Yukawa interactions are described by the parameters $h$, which
are related to the masses of the standard fermions by the usual
expressions, {\em e.g.}, $m_t = h^t v_2, m_b = h^b v_1$, where 
$v_i = <H_i>$. 

 Implementation of  this model within a supergravity scheme 
 leads naturally to a set of unification assumptions at a Grand
 Unification (GUT) scale, $M_{GUT}$:

     i) Unification  of the gaugino masses:
        $M_i(M_{GUT}) \equiv m_{1/2}$,

     ii) Universality of the scalar masses with a common mass denoted by
     $m_0$: $m_i(M_{GUT})$ \hfill \break
    \indent \phantom{ii)\ }  $ \equiv m_0$,

    iii) Universality of the trilinear scalar couplings:
         $A^{l}(M_{GUT}) = A^{d}(M_{GUT}) = A^{u}(M_{GUT})$ \hfill \break
    \indent \phantom{iii)\ }  $\equiv A_0 m_0$. 

 This scheme will be denoted here as universal SUGRA (or simply
 SUGRA). The relevant parameters of the model at the electro--weak 
(EW) scale are obtained from their corresponding values at the 
$M_{GUT}$ scale by running these down according to the renormalization
 group equations (RGE).
 By requiring that  the electroweak symmetry breaking is induced
 radiatively by the soft supersymmetry breaking, one finally reduces
 the model parameters to five: 
$m_{1/2}, m_0, A_0, \tan \beta (\equiv  v_2/v_1)$ and sign $\mu$.
In the present paper, these parameters are varied in the following
ranges: 
$50\;\mbox{GeV} \leq m_{1/2} \leq  1\;\mbox{TeV},\;
m_0 \leq  1\;\mbox{TeV},\;
-3 \leq A_0 \leq +3,\;
1 \leq \tan \beta \leq 50$. 
Notice that a common  upper extreme for the
 mass parameters has been used, and generically set at the value of 1 TeV, as 
a typical scale beyond which the main attractive features of supersymmetry 
fade away. However,  fine-tuning  arguments actually 
set  different bounds  for $m_0$ and $m_{1/2}$
(in universal SUGRA and in nuSUGRA) \cite{bere1}: 
$m_{1/2} \lsim$ hundreds of GeV, whereas $m_0 \lsim $  2-3 TeV. 
In the present paper we did not look specifically into the  
$m_0 \sim 2-3$ TeV window; in Ref. \cite{fmw}
 phenomenology of relic neutralinos in 
this large $m_0$ regime has been analyzed \cite{feng}. 

Models with unification conditions at the GUT scale
represent an  appealing scenario; however,
some of the assumptions listed above, particularly ii) and iii), are not
very solid, since, as was  already emphasized some time ago \cite{com},
universality might occur at a scale higher than $M_{GUT}\sim 10^{16}$
GeV, {\em e.g.}, at the Planck scale. More recently, the possibility that
 the initial scale for the RGE running, $M_I$, might be smaller than 
 $M_{GUT}\sim 10^{16}$ has been raised \cite{gabr,abel}, on the basis of
 a number of string models (see for instance the references quoted in 
\cite{gabr}).  In
Ref. \cite{gabr} it is stressed that $M_I$ might be anywhere between the EW
scale and the Planck scale, with significant consequences  for the size of
the neutralino--nucleon cross section.  

An empirical way of taking into account the uncertainty in $M_I$ 
 consists in allowing deviations in the
unification conditions at $M_{GUT}$.     For instance, deviations from 
universality in the scalar  masses at  $M_{GUT}$, which split 
$M_{H_1}$ from $M_{H_2}$ may be parametrized as

\begin{equation}
M_{H_i}^2 (M_{GUT}) = m_0^2(1 + \delta_i)
\label{eq:dev}
\end{equation}

This is the case of non--universal SUGRA (nuSUGRA) that we considered 
 in Refs. \cite{comp,bere1}, and that we analyse again in this paper. Here the
 parameters  $\delta_i$ which quantify the departure from universality 
for the $M_{H_i}^2$ will be varied in the range (-2,+2). Deviations
 from universality in the Higgs masses have recently been considered
 also in Ref. \cite{el2}. 
Further extensions of deviations from universality in SUGRA models
 which  include squark and/or gaugino masses are discussed, for instance,
 in \cite{acc,cn}. 

   The large uncertainties involved in the choice of the scale $M_I$ 
make the use of SUGRA schemes rather problematic and unpractical: 
the originally appealing feature of a universal SUGRA with few parameters 
fails, because of the need to take into consideration the variability of $M_I$ 
or, alternatively, to add new parameters which quantify the various 
deviation effects from universality at the GUT scale. It
appears  more convenient  to work  with a phenomenological 
susy model whose  parameters are defined directly at the electroweak 
scale. We denote here this effective scheme of MSSM by effMSSM. This   
 provides, at the EW scale, a model, defined in terms of a minimum  number of 
parameters: only those necessary to shape the essentials of the theoretical 
structure of an MSSM, and of its particle content. 
Once all experimental and theoretical constraints are implemented in
this effMSSM model, one may investigate its compatibility with SUGRA
schemes at the desired $M_I$.

In the effMSSM scheme we consider here, we impose  a set of
assumptions at the electroweak scale: 
a) all trilinear parameters are set to zero except those of the third family, 
which are unified to a common value $A$;
b) all squark  soft--mass parameters are taken  
degenerate: $m_{\tilde q_i} \equiv m_{\tilde q}$; 
c) all slepton  soft--mass parameters are taken  
degenerate: $m_{\tilde l_i} \equiv m_{\tilde l}$; 
d) the $U(1)$ and $SU(2)$ gaugino masses, $M_1$ and $M_2$, are 
assumed to be linked by the usual relation 
$M_1= (5/3) \tan^2 \theta_W M_2$ (this is the only GUT--induced
relation we are using, since gaugino mass unification appears to be
better motivated than scalar masses universality). 
As a consequence, the supersymmetric 
parameter space consists of seven independent parameters. We choose them to be: 
$M_2, \mu, \tan\beta, m_A, m_{\tilde q}, m_{\tilde l}, A$ and vary these 
parameters in
the following ranges: $50\;\mbox{GeV} \leq M_2 \leq  1\;\mbox{TeV},\;
50\;\mbox{GeV} \leq |\mu| \leq  1\;\mbox{\rm TeV},\;
80\;\mbox{GeV} \leq m_A \leq  1\;\mbox{TeV},\; 
100\;\mbox{GeV} \leq  m_{\tilde q}, m_{\tilde l} \leq  1\;\mbox{TeV},\;
-3 \leq A \leq +3,\; 1 \leq \tan \beta \leq 50$ ($m_A$ is the mass of
the CP-odd neutral Higgs boson).

The effMSSM scheme proves  very manageable for the susy phenomenology at the 
EW scale; as such, it has been frequently used in the literature in 
connection with relic neutralinos 
(often with the further assumption of slepton--squark mass
degeneracy: 
$m_{\tilde{q}} = m_{\tilde{l}}$) \cite{noi,noiult,kkk,nnn,bg,man}. 
Notice that we are not assuming here slepton--squark mass degeneracy.
In the scatter plots given in this paper only configurations with 
$m_{\tilde{q}} \geq m_{\tilde{l}}$ are shown. 
This mass hierarchy is reminiscent of what is usually obtained
in SUGRA schemes, although in our effMSSM it is not necessarily so.
It is worth reporting that some configurations with inverse hierarchy 
$m_{\tilde{q}} \leq m_{\tilde{l}}$ produce some increase in 
$\xi \sigma_{\rm scalar}^{\rm (nucleon)}$ at low $m_{\chi}$ values (see the
discussion after Figs. 5 in Sect. III).  

We recall that even much larger extensions of the
supersymmetric models could be envisaged: for
instance,   non--unification of the gaugino masses \cite{cn,griest},
and schemes with CP--violating phases \cite{cp}. 
Here we limit our considerations to the schemes previously defined: 
universal SUGRA, nuSUGRA, effMSSM.

 The neutralino is defined 
as the lowest--mass linear superposition of photino ($\tilde \gamma$),
zino ($\tilde Z$) and the two higgsino states
($\tilde H_1^{\circ}$, $\tilde H_2^{\circ}$):
$\chi \equiv a_1 \tilde \gamma + a_2 \tilde Z + a_3 \tilde H_1^{\circ}  
+ a_4 \tilde H_2^{\circ}$. 
Hereafter, the nature of the neutralino is classified in terms of a
parameter $P$, defined as $P \equiv a_1^2 + a_2^2$.  
The neutralino is called a gaugino when $P > 0.9$, a higgsino when 
$P < 0.1$, mixed otherwise. 

For more details concerning theoretical aspects involved in our calculations 
and the way in which the 
experimental constraints due to $b \rightarrow s + \gamma$ is 
 implemented  we refer to Refs. \cite{comp,noiult}. 
 Accelerators data on supersymmetric
and Higgs boson searches (CERN $e^+ e^-$ collider LEP2 and Collider
Detector CDF at Fermilab) provide now rather stringent bounds on
supersymmetric parameters. 
CDF bounds are taken from \cite{cdf}. The new LEP2 bounds are taken
from \cite{LEPb,donan}; these   constrain the 
 configurations of relevance for relic neutralinos more severely 
as compared, for instance, with those considered in Ref. \cite{noiult}.

The results for the neutralino relic abundance have been obtained with 
the procedure indicated in Ref. \cite{noiom}. 
 The neutralino--nucleon cross section has been 
calculated with the formulae reported in Ref. \cite{noi,noi6}. As discussed 
in the introduction, this cross section suffers from significant  
uncertainties in the size of Higgs--quark--quark and 
squark--quark--neutralino couplings. In this paper we use for these 
quantities what we have defined as set 1 and set 2 in Ref. \cite{noi6} 
to which we refer for details. Here we only report the  values of the 
quantities   $m_{q}<\bar{q}q>$ for the two sets:

\begin{eqnarray} \label{eq:set1}
 \mbox{Set 1:}\;
 m_{l}<\bar{l}l>\; =\; 23\; {\rm MeV}, \;
  m_{s}<\bar{s}s>\; =\; 215\; {\rm MeV}, \;
  m_{h}<\bar{h}h>\; =\; 50\; {\rm MeV}. 
\end{eqnarray} 

\medskip
\begin{eqnarray} \label{eq:set2}
 \mbox{Set 2:}\;
 m_{l}<\bar{l}l>\; =\; 30\; {\rm MeV}, \;
  m_{s}<\bar{s}s>\; =\; 435\; {\rm MeV}, \;
  m_{h}<\bar{h}h>\; =\; 33\; {\rm MeV}.
\end{eqnarray}

\noindent
In Eqs. (\ref{eq:set1}-- \ref{eq:set2}) $l$ stands for light quarks, $s$ is
the strange quark and $h=c,b,t$ denotes heavy quarks. For the light
quarks, we have defined 
$m_{l}<\bar l l>$ $\equiv$ 
$\frac{1}{2}[m_u <\bar u u> + m_d <\bar d d>]$.
\noindent
Set 1 and set 2 bracket, only partially, the present uncertainties. 
In Ref. \cite{noi6} we also considered 
the consequences of using a more extreme set of values.
 It is worth  recalling that the quantity 
$m_{s}<\bar{s}s>$ is crucial in establishing the size of 
$\sigma_{\rm scalar}^{(\rm nucleon)}$ \cite{ggr}.

The results  shown in the next section are obtained with the same numerical
codes employed in our previous papers
\cite{noi,comp,noi5,noiult,noi6}, but take into account all new 
accelerator data.

\section{Results}

We turn now to the presentation of our results. In Figs. 1a-c we
give the scatter plots for $\sigma_{\rm scalar}^{(\rm nucleon)}$ versus 
$\Omega_{\chi} h^2$ for the three different schemes: universal SUGRA, 
non--universal SUGRA and effMSSM. 
For the SUGRA schemes we only display the results corresponding to positive 
values of $\mu$, since, for  negative values,
the constraint on $b \rightarrow s + \gamma$ implies a large suppression of 
$\sigma_{\rm scalar}^{(\rm nucleon)}$. 
 The two horizontal lines bracket
the sensitivity region defined by Eq. (\ref{eq:section}). The two
vertical lines denote a favorite range for $\Omega_{m}h^2$, 
$0.05 \leq \Omega_{m} h^2 \leq 0.3$, as derived from a host of
 observational data.  According to the most recent determinations
 \cite{fr},  
the lower bound on $\Omega_{m} h^2$ is approaching the value 0.08. 
However, due to the still unsettled 
situation as regards determinations of the matter density in the universe
and of the Hubble constant, a conservative attitude seems advisable.
Anyway, we stress that in the present paper we are not restricting
ourselves
 to any particular 
interval of $\Omega_{m} h^2$. Only some features of Figs. 5 
depend on the actual value employed for the minimum amount of matter 
necessary to reproduce the halo  properties correctly.

Figs. 1a-c provide a first relevant result of our analysis: the
present experimental sensitivity in WIMP direct searches allows the
exploration of supersymmetric configurations of cosmological interest,
also in the constrained SUGRA scheme. It is remarkable that the upper
frontier of the scatter plots is not significantly different in the
three different models, although the region of experimental
sensitivity and cosmological interest is covered with an increasingly
larger variety of supersymmetric configurations as one moves from
SUGRA to nuSUGRA and to effMSSM. This latter fact is expected from the
intrinsic features of the various schemes.  This point will be further
discussed later on, in connection with Figs. 4.  Fig. 2 shows what is
the effect of using set 2 instead of set 1 for the quantities
$m_{q}<\bar{q}q>$'s in effMSSM.  

Once a measurement of the quantity $\rho_{\chi} \cdot \sigma^{\rm
  (nucleon)}_{\rm scalar}$ is performed, values for the local density
$\rho_{\chi}$ versus the relic abundance $\Omega_{\chi}h^2$ may be deduced by
proceeding in the following way \cite{noi6}:

\noindent
1)  $\rho_{\chi}$ is evaluated as 
$[\rho_{\chi} \cdot \sigma^{\rm (nucleon)}_{\rm scalar}]_{expt}$ /
$\sigma^{\rm (nucleon)}_{\rm scalar}$, 
where $[\rho_{\chi} \cdot \sigma^{\rm (nucleon)}_{\rm scalar}]_{expt}$ 
denotes the experimental value, and 
$\sigma^{\rm (nucleon)}_{\rm  scalar}$ is calculated as indicated above;
2) to each value of  $\rho_{\chi}$ one associates the corresponding
calculated value of $\Omega_{\chi} h^2$. 
The scatter plot in Fig. 3 is derived from the lowest value of the
annual--modulation region of Ref. \cite{damalast},  
$[\rho_{\chi}$/(0.3 GeV cm$^{-3}$) $\cdot \sigma^{\rm (nucleon)}_{\rm
    scalar}]_{expt} = 1 \cdot 10^{-9}$ nbarn, and by taking 
$m_{\chi}$ in the range of Eq. (\ref{eq:mass}). 
 This  plot,
obtained in case of effMSSM, shows that the most interesting region,  
{\it i. e.} the one 
with  0.2 GeV cm$^{-3} \leq \rho_{\chi} \leq $ 0.7 GeV cm$^{-3}$ 
 and $0.05 \leq \Omega_{m} h^2 \leq 0.3$ (cross-hatched region in 
the figure), is covered
   by susy configurations probed by the WIMP direct detection.

Let us examine the various sectors of Fig. 3. 
 Configurations above the upper horizontal line are
incompatible with the upper limit on the local density of dark
matter in our Galaxy and must be disregarded.
Configurations above the 
upper slanted dot--dashed line and below the upper horizontal solid line 
would imply a stronger clustering of neutralinos in our halo as 
compared to their average distribution in the Universe. This
situation may be considered unlikely, since in this case
neutralinos could fulfill the experimental range for 
$\rho_\chi$, but they would contribute only a small fraction to
the cosmological cold dark matter content.
For configurations which fall inside 
the band delimited by the slanted dot--dashed lines and simply--hatched 
in the figure,
the neutralino would provide only a fraction of the cold dark 
matter at the level of local density and of the 
average relic abundance, a situation which would be possible, for instance,
if the neutralino is not the unique cold dark matter particle
component. To neutralinos belonging to these configurations one 
should assign a {\it rescaled} local density 
$\rho_{\chi} = \rho_l \times \Omega_{\chi} h^2/(\Omega_m h^2)_{min}$, where 
$(\Omega_m h^2)_{min}$ is the minimum value of $\Omega_m h^2$ 
compatible with halo properties.

It is interesting to analyze the properties pertaining to the
supersymmetric configurations which stay inside the favored region 0.2
GeV cm$^{-3} \leq \rho_{\chi} \leq $ 0.7 GeV cm$^{-3}$ and $0.05 \leq
\Omega_{m} h^2 \leq 0.3$ and to those which stay inside the corridor
where rescaling applies ({\it i.e.} the corridor between the two
slanted dot--dashed lines). In Figs. 4a-c we give the scatter plots
for all these configurations in the plane $m_h - \tan \beta$ ($m_h$
being the mass of the lightest CP--even neutral Higgs boson).

Let us make a few comments about these results.  First, we note that a
feature that was already pointed out in Ref. \cite{comp} is recovered:
in SUGRA (Fig. 4a) only high values of $\tan \beta$, $\tan \beta \gsim
40$, are involved in present direct detection experiments. Similar
conclusions are also reached in more recent papers \cite{acc,lns}.
The occurrence of the lower bound $\tan\beta \gsim 40$ in the SUGRA
scheme is a consequence of RGE evolutions of the parameters and the
nature of radiative electroweak symmetry breaking, which induce strong
correlations among the parameters at the low energy scale
\cite{bere1}. In this class of models, couplings of the light Higgs
boson $h$ to the $s$ quark can be enhanced only for large values of
$\tan\beta$. As a consequence, the neutralino-nucleon cross section
can be substantially large only when $\tan \beta \gsim 40$.
Configurations displayed in Fig. 4a entail relatively light
pseudoscalar Higgs $A$, $m_A \lsim 200$ GeV, $m_0 \gsim 350$ GeV, and
lightest stop and sbottom masses larger than about $460$ GeV.
   
These features are somewhat (even though not completely) relaxed in
the nuSUGRA scheme, where the non--universality in the Higgs sector
allows for milder correlations among the parameters and among the low
energy variables. This is especially true for the lower bound on
$\tan\beta$, which now moves down to about 7, as displayed in Fig. 4b.
Also for the other parameters we have weaker bounds with respect to
the SUGRA case. For instance: $m_A \lsim 450$ GeV, $m_0 \gsim 200$
GeV, and lightest stop and sbottom masses larger than about 150 GeV
and 400 GeV, respectively.

In the effMSSM scheme most of the internal correlations of the model,
which are typical of the supergravity-inspired schemes, are not
present. For instance, $\tan\beta$ and $m_A$ are now independent
parameters. Also in this case, the most relevant information about
configurations which give $\rho_\chi$ and $\Omega_{m} h^2$ inside the
favoured region defined above is provided by the $\tan\beta$--$m_h$
correlation, which is shown in Fig. 4c.  We notice that in effMSSM the
lower bound on $\tan\beta$ is around 5. In Fig. 4c we also display, by
a dashed line, what would be the boundary of the scatter plot, in case
set 2 for the quantities $m_{q}<\bar{q}q>$'s is employed instead of
set 1. As for other relevant correlations, we find, for instance: $m_A
\lsim 500$ GeV and/or $m_{\tilde q} \lsim 300$ GeV.

As a final comment, we point out that in our calculations we have
taken into account the experimental constraint on $\sin^2(\alpha -
\beta)$ \cite{donan}. This limit is stronger than the one displayed in
terms of $\tan \beta$ versus $m_h$ in Figs. 4a-c and has the effect of
depopulating the scatter plots, without modifying their boundaries.  
 
We wish to point out that, should the continuation of the LEP running 
provide some support in favor of  a Higgs boson at a mass of about  115 GeV
\cite{higgs}, this would fit remarkably well  within the
configurations displayed in Figs. 4 \cite{el3}. It would be a case where
accelerator measurements and WIMP searches would complement each other 
in  providing a possible, consistent picture of physics beyond the
standard model.    

For sake of comparison with specific experimental results, we provide in Figs.
5a-c the scatter plots for the quantity $\xi \sigma^{\rm (nucleon)}_{\rm
  scalar}$ versus $m_{\chi}$ in the various supersymmetric schemes.  $\xi$ is
taken to be $\xi = {\rm min}\{1, \Omega_{\chi} h^2/(\Omega_m
  h^2)_{min}\}$, in order to have rescaling in the neutralino local density,
when $\Omega{\chi} h^2$ turns out to be less than $(\Omega_m h^2)_{min}$ (here
$(\Omega_m h^2)_{min}$ is set to the value 0.05).  In universal SUGRA our
results reach a maximum for $\xi \sigma_{\rm scalar}^{(\rm nucleon)}$ at the
level of about $10^{-9}$ nb, a feature which is in common, for instance, with
the results of Refs. \cite{acc,cn,man}. Lower values for the WIMP--nucleon
cross section are found in evaluations where various inputs, each one having
the effect of suppressing the value of $\xi \sigma^{\rm (nucleon)}_{\rm
  scalar}$, are employed concomitantly \cite{efo,el2}: i) low values for $\tan
\beta$, $\tan \beta \leq 10$; ii) small values for the quantity
$m_{s}<\bar{s}s>$, iii) a tight lower bound on the neutralino relic abundance
$\Omega_{\chi} h^2 > 0.1$ . In fact, should we use the same inputs as in
\cite{efo}, we would obtain the scatter plot which stays below the dashed line
displayed in Fig. 5a, in agreement with the results of Ref. \cite{efo}.

However, we point out that, in general, 
 in universal SUGRA, evaluations by various authors  differ in some
  features, for instance in the position of the maximum of
  $\xi \sigma_{\rm scalar}^{(\rm nucleon)}$ in terms of $m_{\chi}$. 
 This is likely to be due to the fact that this
 strict scheme is very sensitive to the specific ways in which
 various constraints (for instance, b $\rightarrow$ s + $\gamma$) are
 implemented in the calculations.

In Figs. 5a-c the solid line denotes the frontier of the 3$\sigma$ 
annual--modulation region of Ref. \cite{damalast}, when only the 
uncertainties in $\rho_l$ and in the dispersion velocity of a
Maxwell--Boltzmann distribution, but  not the ones in other
astrophysical quantities, 
 are taken into account. As discussed in the
Introduction,  effects due to a possible bulk rotation of the dark halo or to
an asymmetry in the WIMP velocity distribution would move this
boundary towards 
higher values of $m_{\chi}$. 
Our results in Figs. 5a-c show that the susy scatter plots reach up 
the annual--modulation region of  
Ref. \cite{damalast}, even with the current stringent bounds from
accelerators (obviously, more easily in effMSSM than in SUGRA and
nuSUGRA schemes).

In connection with the results shown in Fig. 5c for effMSSM, we
further  remark that, if  configurations with the hierarchy 
$m_{\tilde{q}} < m_{\tilde{l}}$ are included, the scatter plot would rise 
by a factor of a few  at $m_{\chi} \sim 50-90$ GeV.

Finally, we recall that use of set 2 for the quantities 
 $m_{q}<\bar{q}q>$'s instead of set 1 would entail an increase of
 about a factor 3 in all the scatter plots of Figs. 5.

\section{Conclusions}

In this work we have shown that the current direct experiments for
WIMPs, when interpreted in terms of relic neutralinos, are indeed
probing regions of the supersymmetric parameter space compatible with
all present bounds from accelerators. We have quantified the extent of
the exploration attainable by WIMP direct experiments in terms of
various supersymmetric schemes, from a SUGRA scheme with unification
assumptions at the grand unification scale to an effective model,
effMSSM, at the electroweak scale. It has been stressed that, due the
large uncertainties in the unification assumptions in SUGRA schemes,
the effMSSM framework turns out to be the most convenient model for
neutralino phenomenology.
 
We have proved that part of the configurations probed by WIMP
experiments, and not disallowed by present accelerator bounds, entail
relic neutralinos of cosmological interest. As discussed in the
previous Section, this result is at variance with the conclusions of
some analyses recently appeared in the literature.  Also neutralinos
which might contribute only partially to the required amount of dark
matter in the universe have been included in our analysis. The
cosmological properties have been displayed in terms of a plot of the
local density versus the average relic abundance, {\it i.e.} in a
representation which proves particularly useful to summarize the
properties of relic neutralinos.

We have noticed that a Higgs with a mass of about 115 GeV, such as the
one now under experimental investigation at LEP2, would fit remarkably
well in the above scenario.

In our evaluations we have taken into account that the determination
of the actual sensitivity region in terms of the WIMP--nucleon cross
section and of the WIMP mass from the experimental data depends quite
sizeably on uncertainties of various origins, mainly: i) possible
effects due to a halo bulk rotation and/or to asymmetries in the WIMP
velocities distribution, ii) significant uncertainties in the
determination of Higgs--quark--quark and neutralino--quark--squark
couplings. We have stressed that all these effects have to be taken
properly into account, when conclusions about comparison of theory
with experiments are drawn.

\acknowledgements 
This work was partially supported by the Research
Grants of the Italian Ministero dell'Universit\`a e della Ricerca
Scientifica e Tecnologica (MURST) within the {\sl Astroparticle
  Physics Project}.  N.F. wishes to thank the kind hospitality of
LAPTH in Annecy where part of this work has been done.

\newpage
\begin{figure}[t]
\hbox{
\psfig{figure=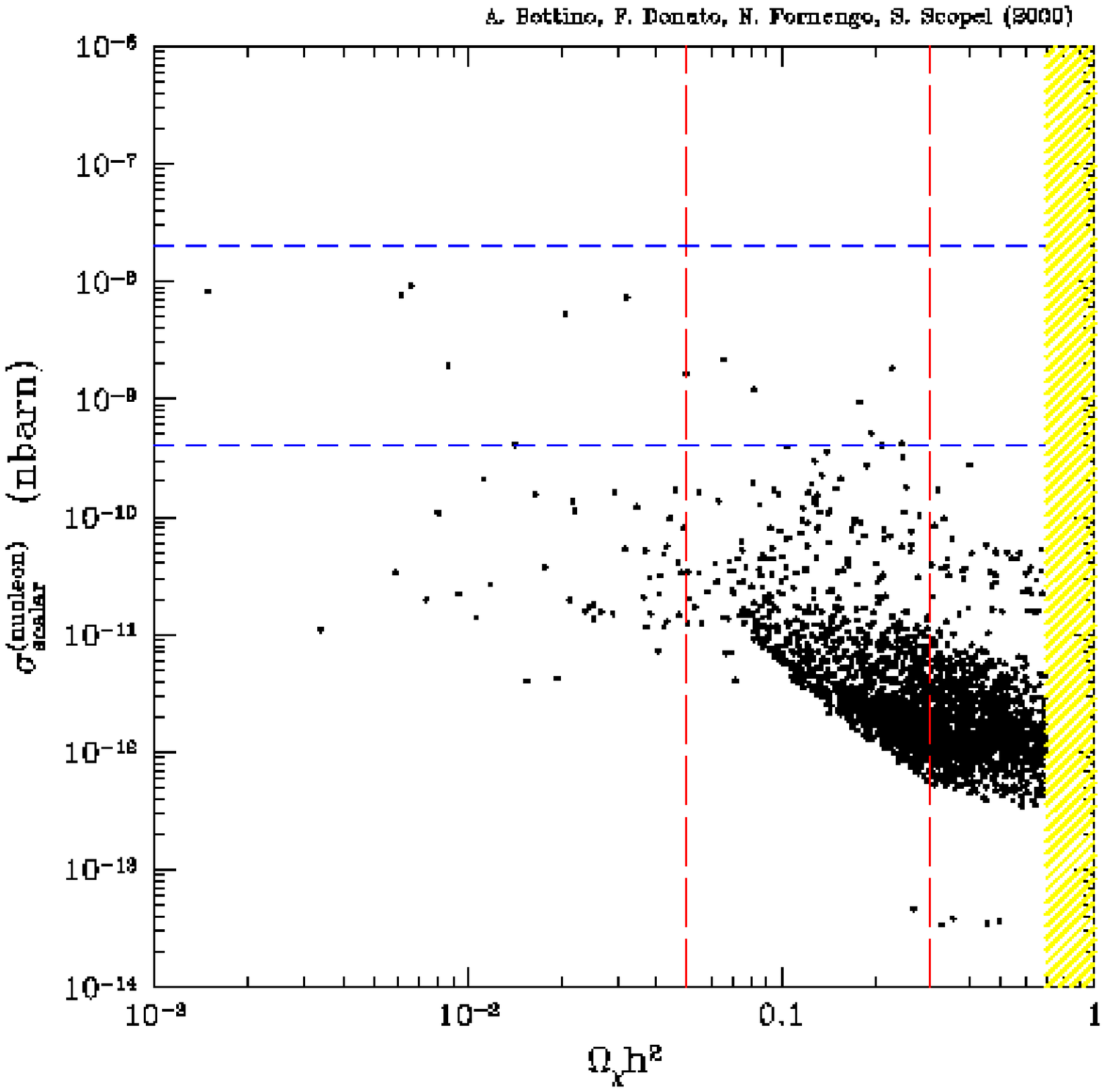,width=8.2in,bbllx=40bp,bblly=180bp,bburx=700bp,bbury=660bp,clip=}
}
{
FIG. 1a.
Scatter plot of $\sigma_{\rm scalar}^{(\rm nucleon)}$ versus 
$\Omega_{\chi} h^2$ for  universal SUGRA. Set 1 for the 
quantities $m_{q}<\bar{q}q>$'s is employed.
Only configurations with positive $\mu$ are shown and 
$m_{\chi}$ is taken in the range of Eq. (\ref{eq:mass}).
 The two horizontal lines bracket the sensitivity region defined 
 by Eq. (\ref{eq:section}). 
 The two vertical lines denote the range 
$0.05 \leq \Omega_{m} h^2 \leq 0.3$.
The region above $\Omega_{\chi} h^2 = 0.7$ is excluded by current limits on
the age of the universe. 
 All points of this scatter plot 
denote gaugino configurations. 
}
\end{figure}

\newpage
\begin{figure}[t]
\hbox{
\psfig{figure=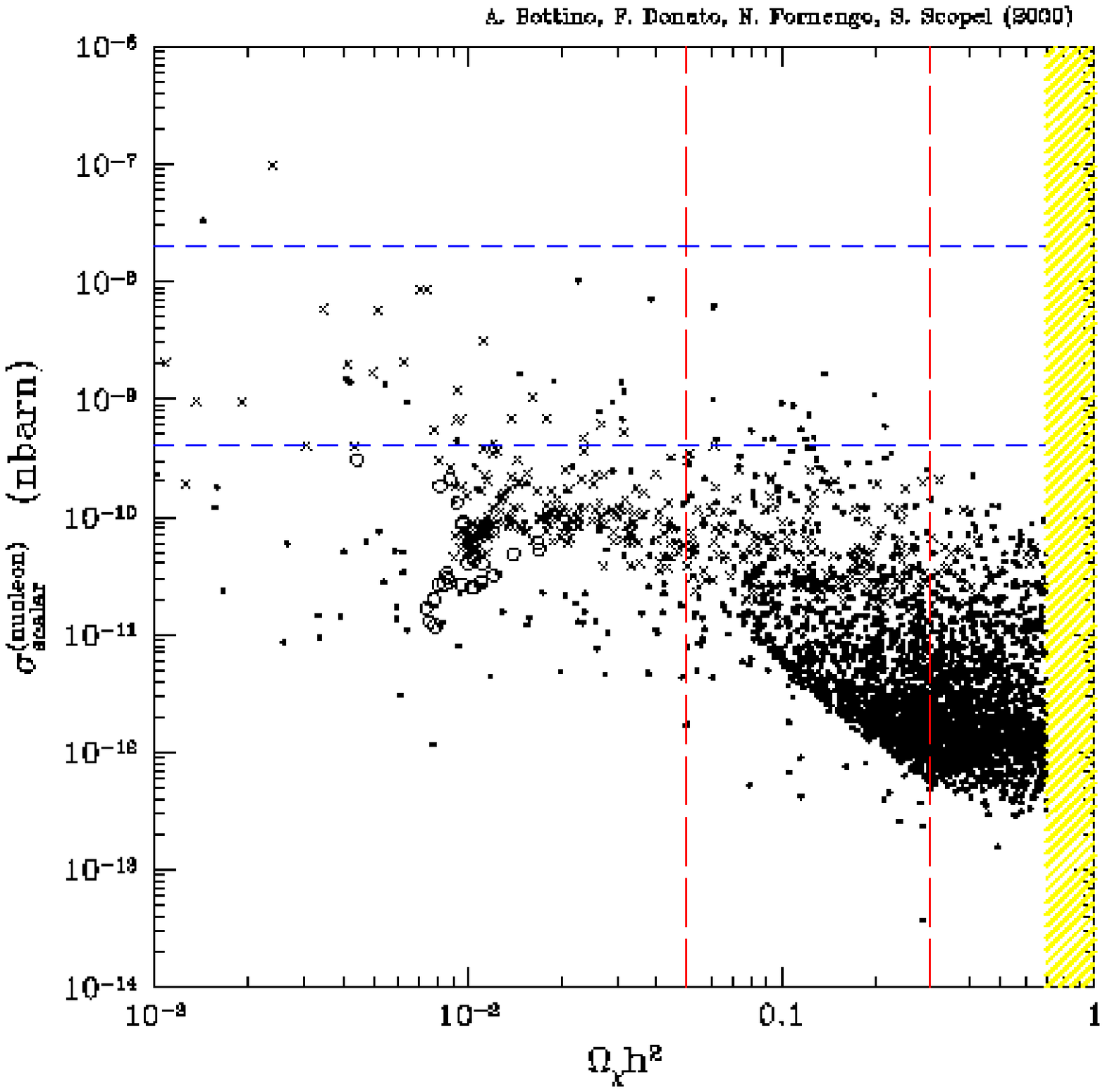,width=8.2in,bbllx=40bp,bblly=180bp,bburx=700bp,bbury=660bp,clip=}
}
{
FIG. 1b.
Scatter plot of $\sigma_{\rm scalar}^{(\rm nucleon)}$ versus 
$\Omega_{\chi} h^2$ for   nuSUGRA. Notations as in Fig. 1a, except that 
here the scatter plot contains neutralinos of various configurations: 
dots denote gauginos, circles denote higgsinos and crosses denote mixed 
configurations. 
}
\end{figure}

\newpage
\begin{figure}[t]
\hbox{
\psfig{figure=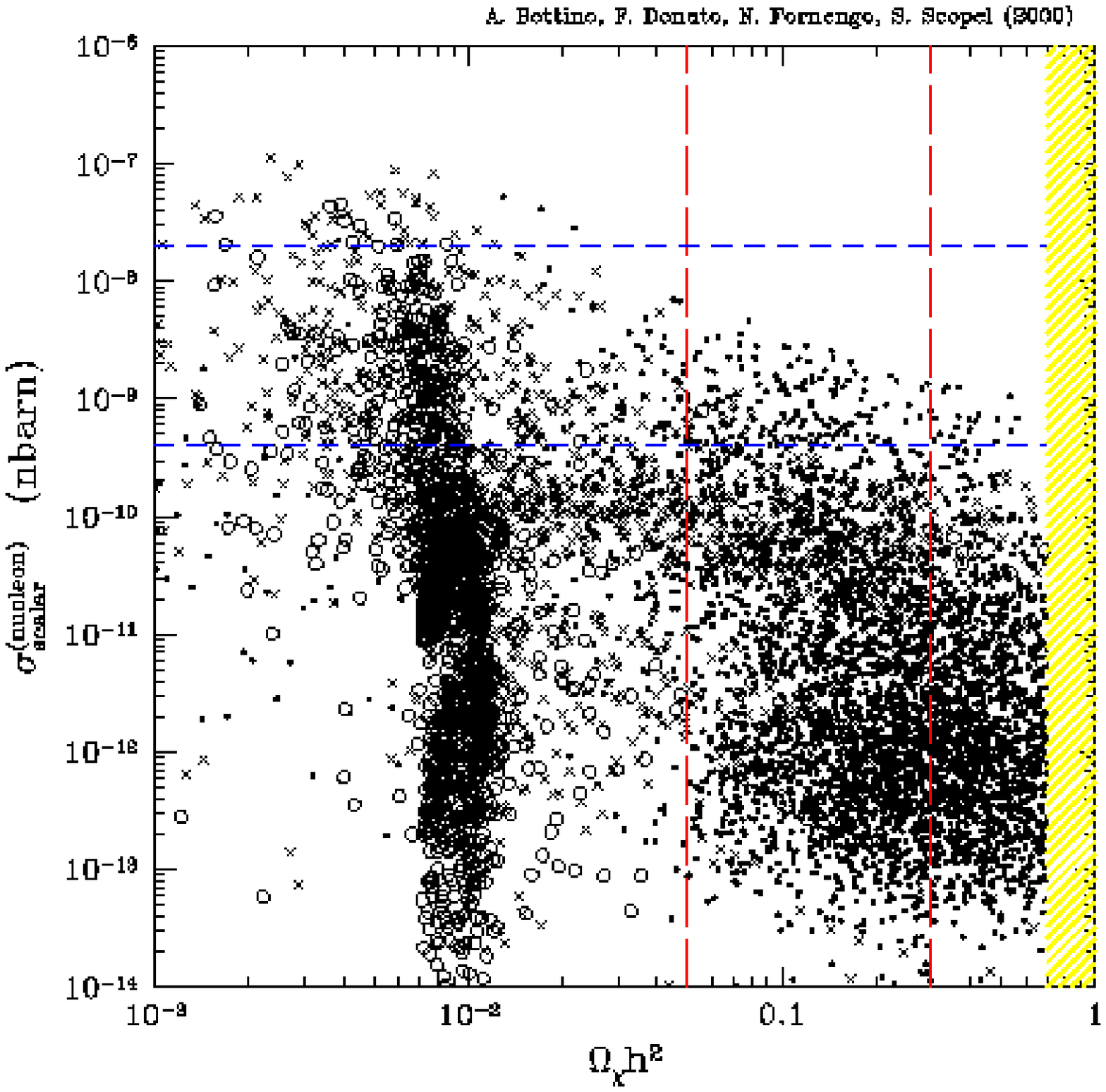,width=8.2in,bbllx=40bp,bblly=180bp,bburx=700bp,bbury=660bp,clip=}
}
{
FIG. 1c.
Scatter plot of $\sigma_{\rm scalar}^{(\rm nucleon)}$ versus 
$\Omega_{\chi} h^2$ for  effMSSM. Notations as in Fig. 1b. Both signs
of $\mu$ are shown. 
}
\end{figure}

\newpage
\begin{figure}[t]
\hbox{
\psfig{figure=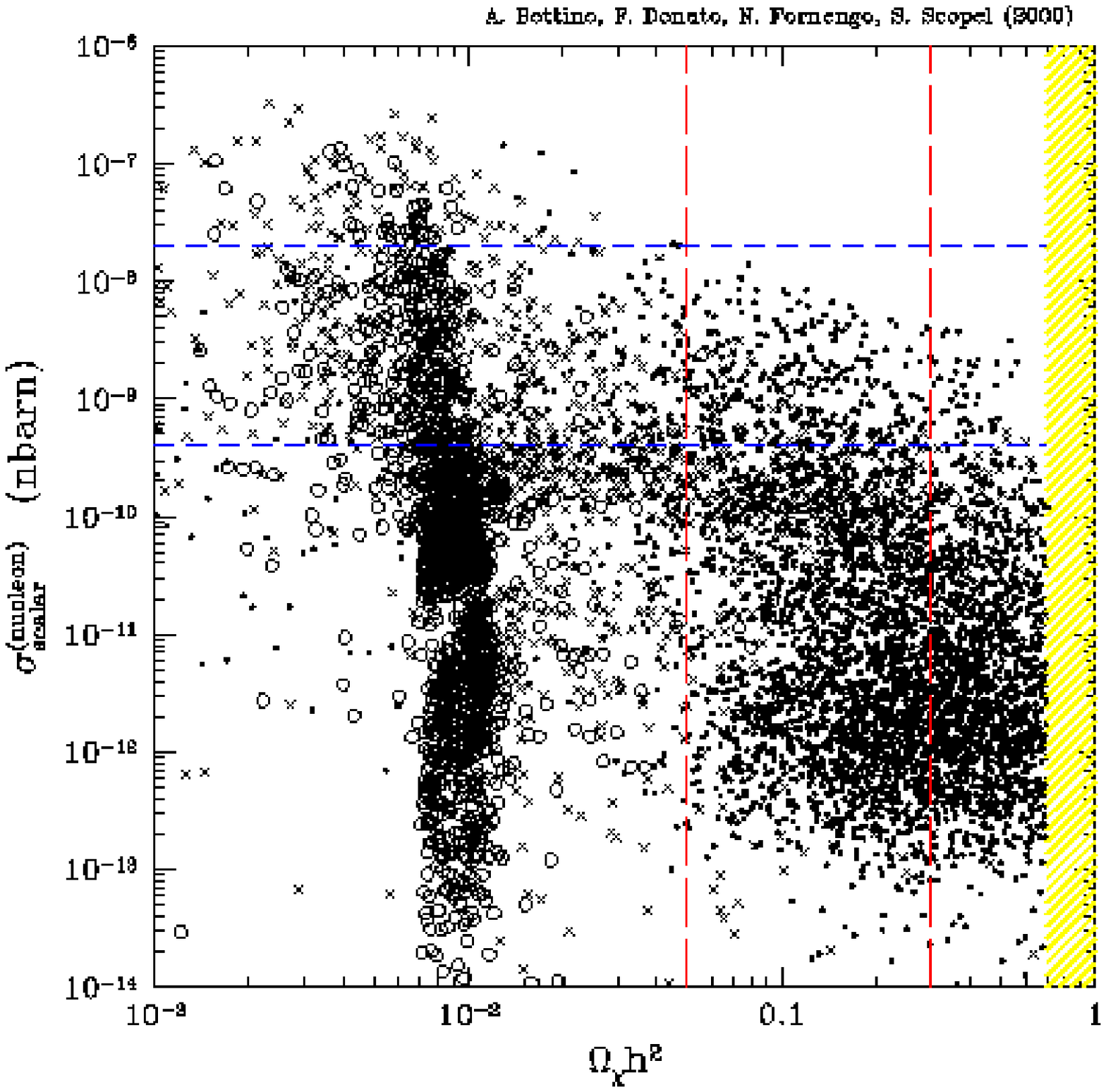,width=8.2in,bbllx=40bp,bblly=180bp,bburx=700bp,bbury=660bp,clip=}
}
{
FIG. 2.
Same as in Fig. 1c, except that here 
set 2 for the quantities $m_{q}<\bar{q}q>$'s is employed instead of
set 1. 
}
\end{figure}

\newpage
\begin{figure}[t]
\hbox{
\psfig{figure=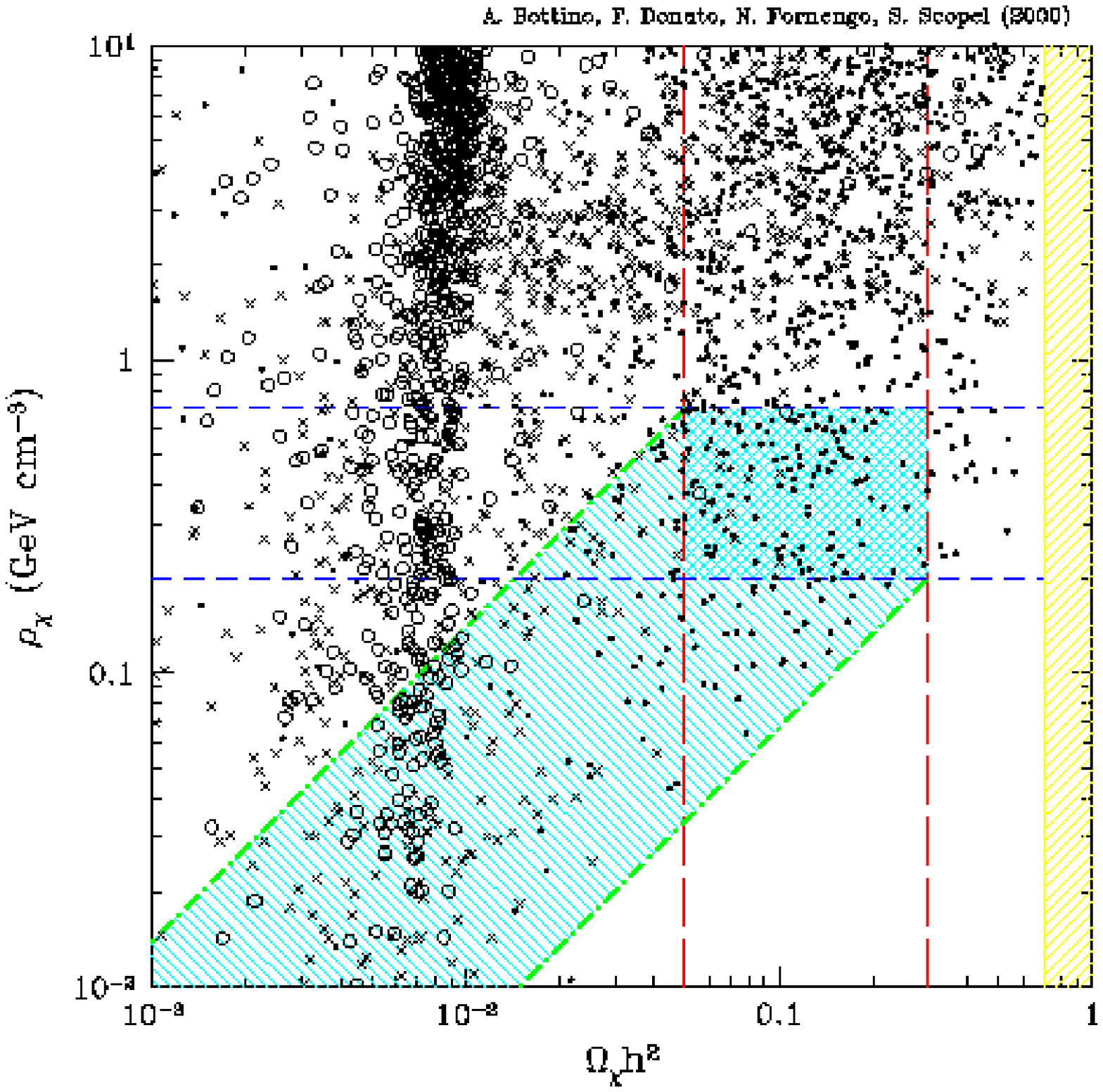,width=8.2in,bbllx=40bp,bblly=180bp,bburx=700bp,bbury=660bp,clip=}
}
{
FIG. 3.
Scatter plot of 
$\rho_{\chi}$ versus $\Omega_{\chi}h^2$. This plot is derived 
from the experimental value  
$[\rho_{\chi}$/(0.3 GeV cm$^{-3}$) $\cdot \sigma^{\rm (nucleon)}_{\rm
    scalar}]_{expt} = 1 \cdot 10^{-9}$ nbarn and by taking 
$m_{\chi}$ in the range of Eq. (\ref{eq:mass}), 
  according to the  procedure outlined in the text, 
in case of effMSSM. 
Set 1 for the quantities $m_{q}<\bar{q}q>$'s is employed.
The two horizontal lines delimit the range 
 0.2 GeV cm$^{-3} \leq \rho_{\chi} \leq $ 0.7 GeV cm$^{-3}$; 
the two vertical ones delimit the range 
  $0.05 \leq \Omega_{m} h^2 \leq 0.3$. 
The region above $\Omega_{\chi} h^2 = 0.7$ is excluded by current limits on
the age of the universe.  
The band delimited by the two slanted dot--dashed lines and simply hatched
is the region where rescaling of $\rho_l$ applies. 
Dots denote gauginos, circles denote higgsinos and crosses denote mixed 
configurations. 
}
\end{figure}

\newpage
\begin{figure}[t]
\hbox{
\psfig{figure=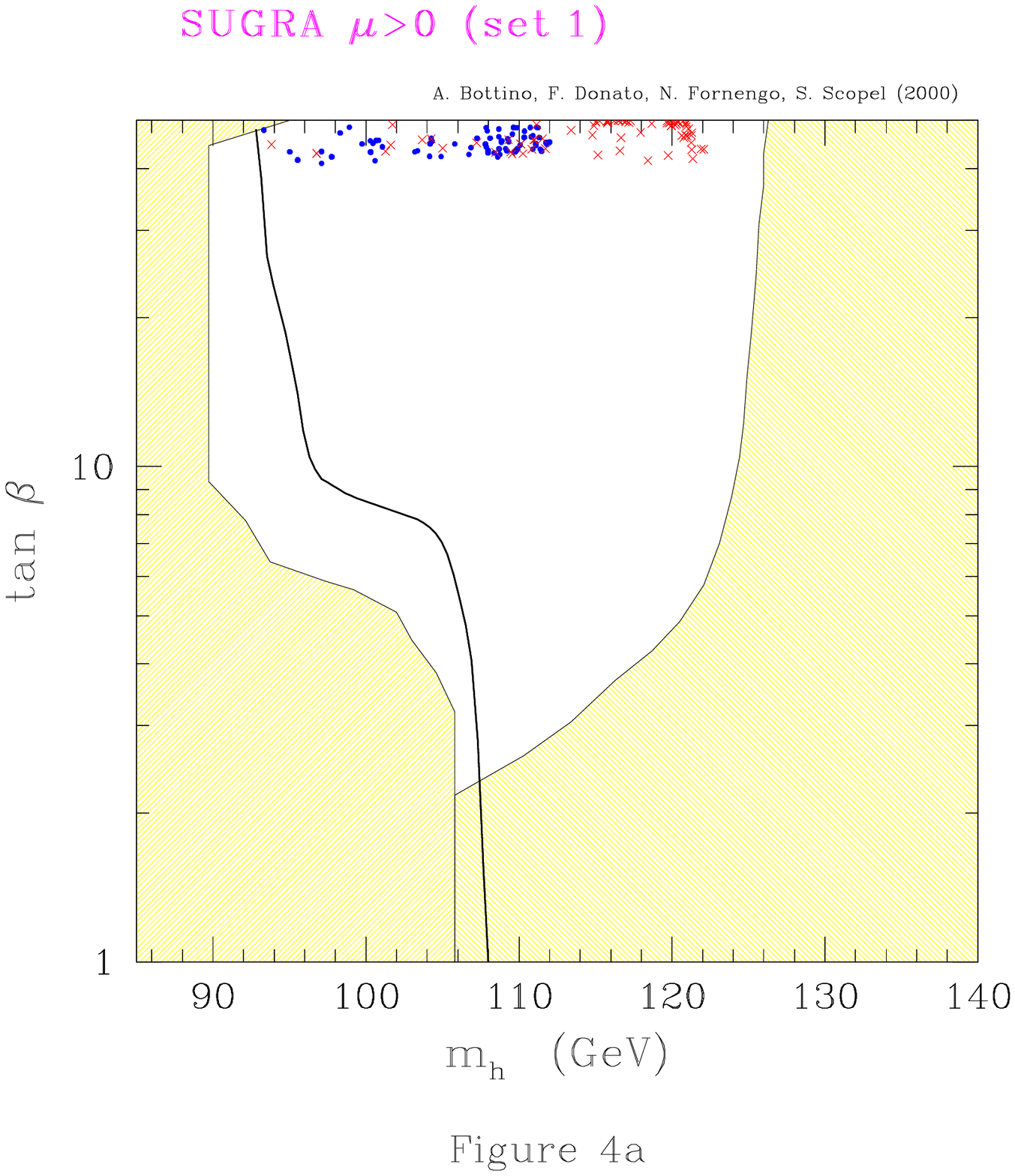,width=8.2in,bbllx=40bp,bblly=180bp,bburx=700bp,bbury=660bp,clip=}
}
{
FIG. 4a.
Scatter plot  in the plane $m_h - \tan \beta$ of the SUGRA 
supersymmetric configurations which stay either inside the  region 
  0.2 GeV cm$^{-3} \leq \rho_{\chi} \leq $ 0.7 GeV cm$^{-3}$ 
 and $0.05 \leq \Omega_{m} h^2 \leq 0.3$ or  
within the  corridor of rescaling in the plane 
$\rho_{\chi}$ versus $\Omega_{\chi}h^2$.  
Set 1 for the quantities $m_{q}<\bar{q}q>$'s is employed.
Crosses (dots) denote configurations with $\Omega_{\chi} h^2 > 0.05$ 
($\Omega_{\chi} h^2 < 0.05$). 
The hatched region on the right is excluded by theory. 
The hatched region on the left is 
excluded by present data from LEP \cite{LEPb} and CDF \cite{cdf}. 
 The solid line represents the 
95\% C.L. bound reachable at LEP2, in case of non discovery of a neutral 
Higgs boson. 

}
\end{figure}

\newpage
\begin{figure}[t]
\hbox{
\psfig{figure=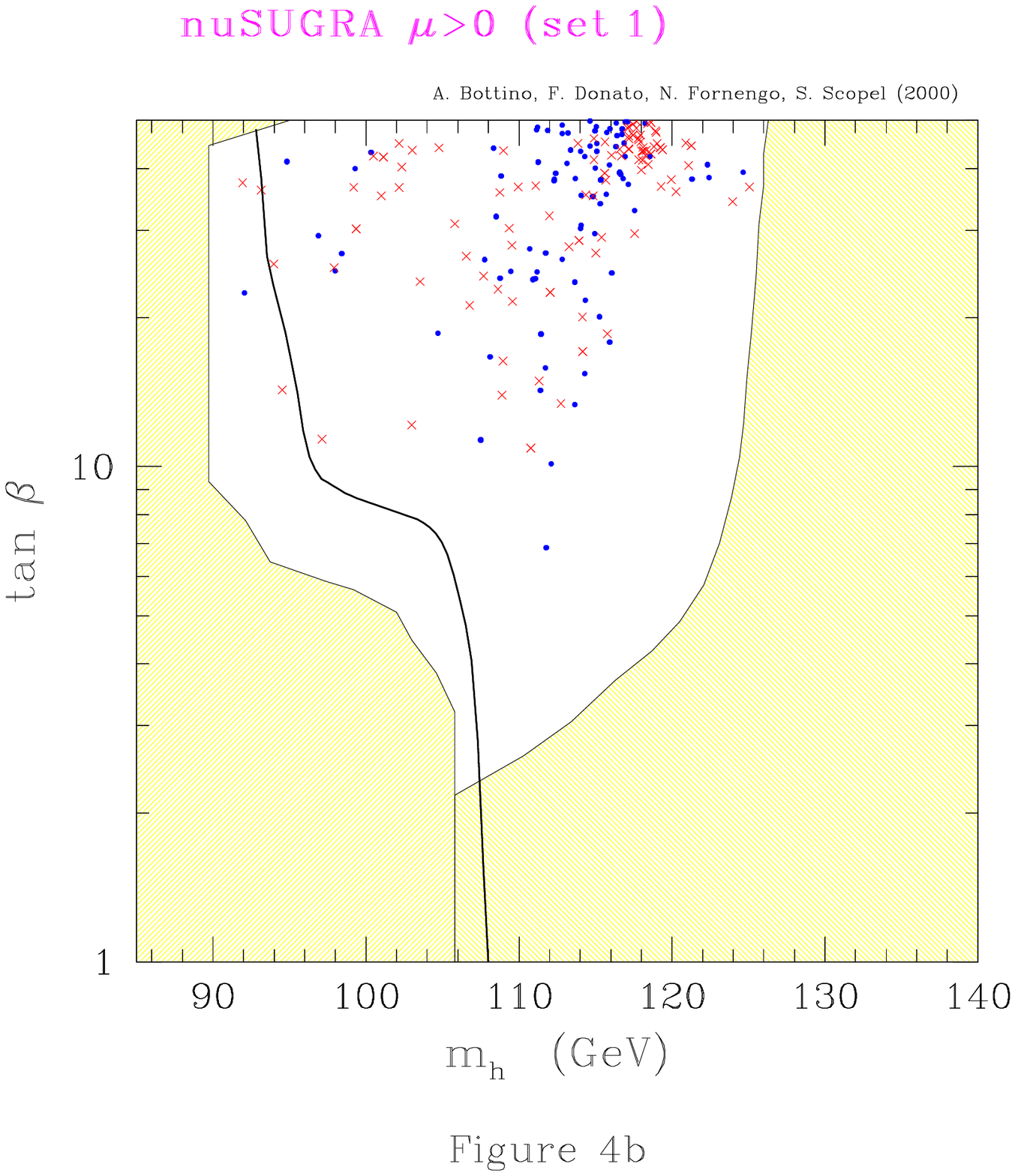,width=8.2in,bbllx=40bp,bblly=180bp,bburx=700bp,bbury=660bp,clip=}
}
{
FIG. 4b.
Same as in Fig. 4a for configurations in  nuSUGRA.
}
\end{figure}

\newpage
\begin{figure}[t]
\hbox{
\psfig{figure=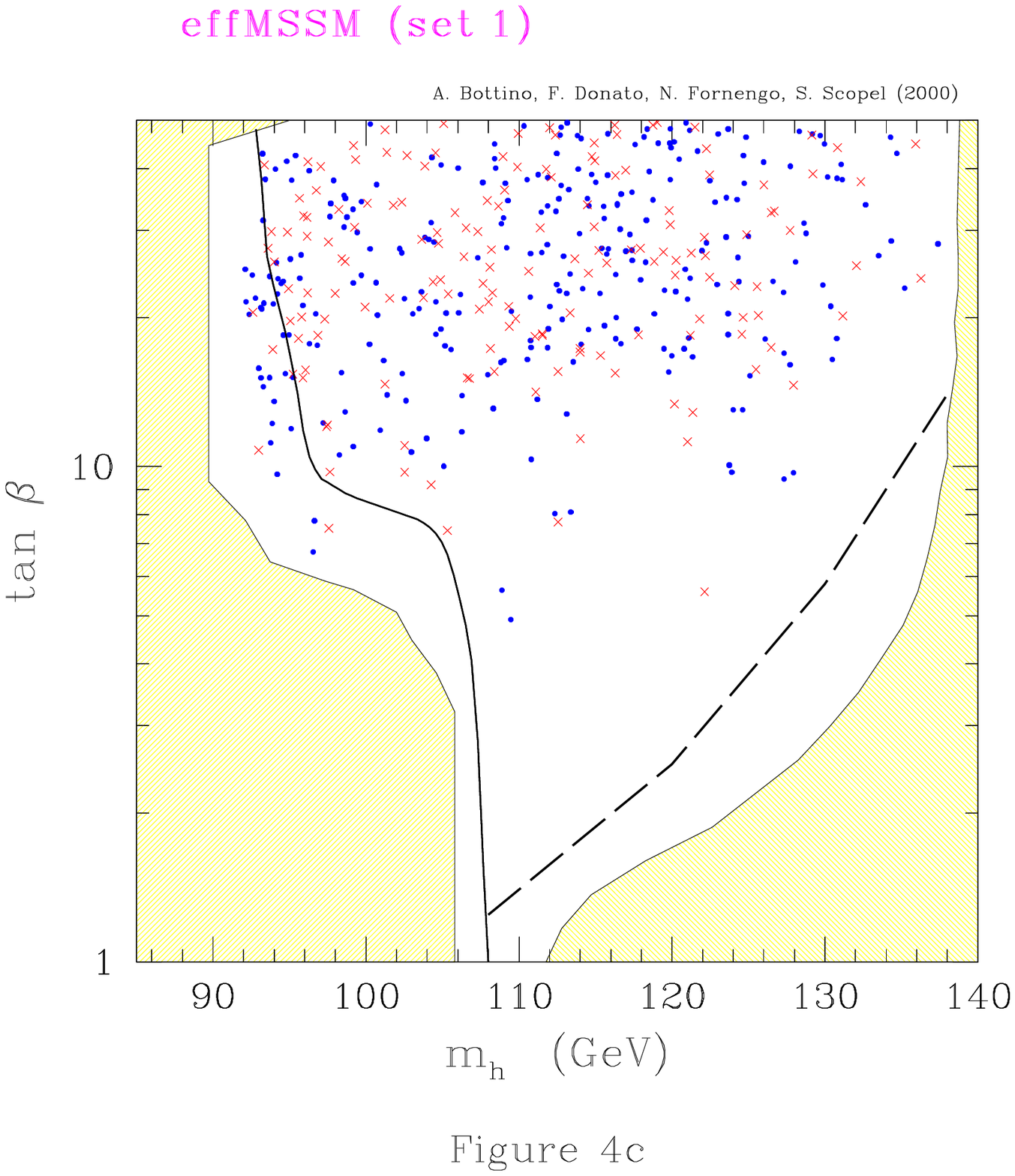,width=8.2in,bbllx=40bp,bblly=180bp,bburx=700bp,bbury=660bp,clip=}
}
{
FIG. 4c.
Same as in Fig. 4a for configurations in effMSSM. The dashed line
denotes to which extent the scatter plot expands if set 2 for the
quantities $m_{q}<\bar{q}q>$'s is used.
}
\end{figure}

\newpage
\begin{figure}[t]
\hbox{
\psfig{figure=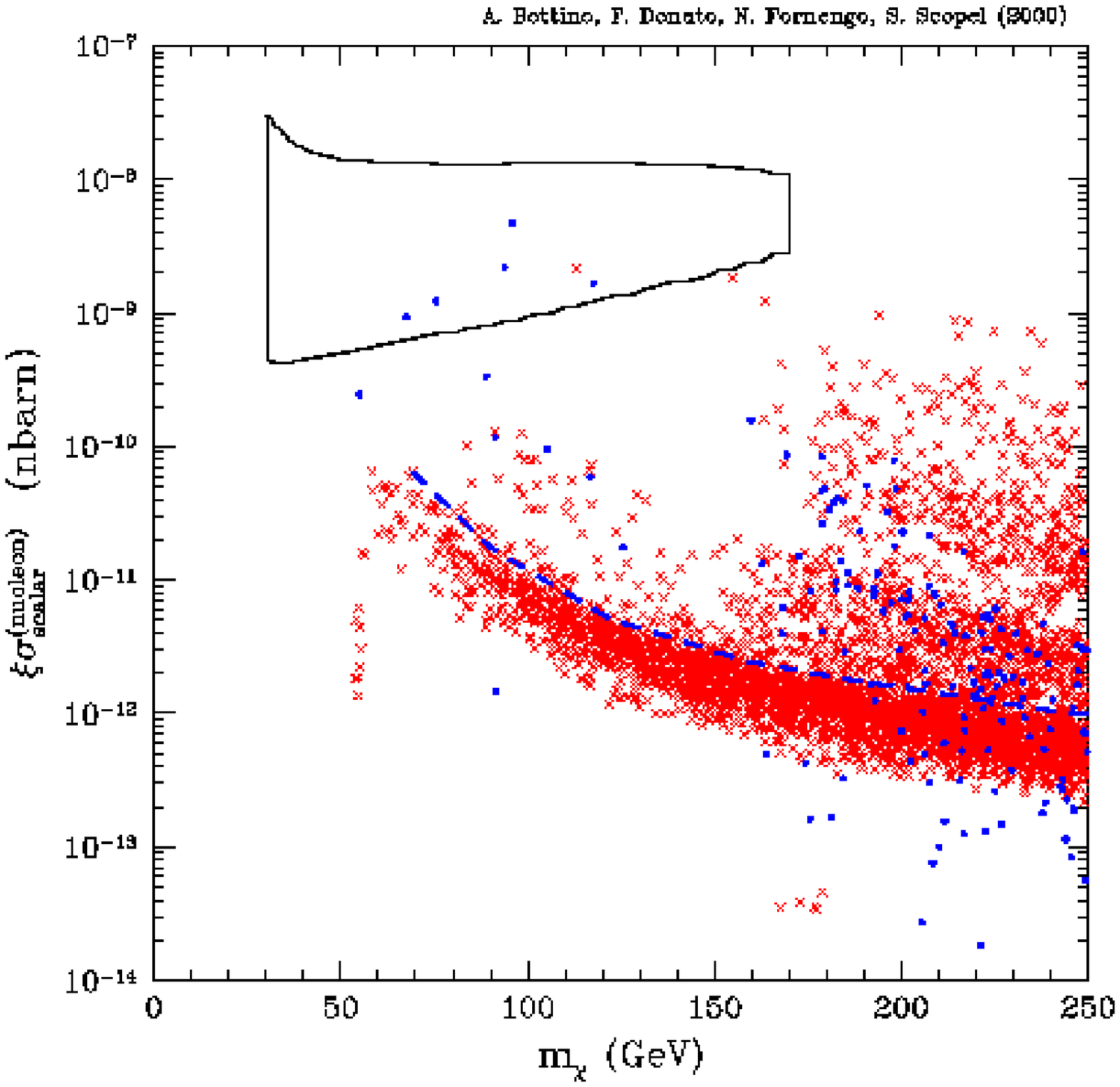,width=8.2in,bbllx=40bp,bblly=180bp,bburx=700bp,bbury=660bp,clip=}
}
{
FIG. 5a.
Scatter plot of $\xi \sigma_{\rm scalar}^{(\rm nucleon)}$ versus 
$m_{\chi}$ in case of universal SUGRA. 
Set 1 for the quantities $m_{q}<\bar{q}q>$'s is employed.
Crosses (dots) denote 
configurations with $\Omega_{\chi} h^2 > 0.05$ 
($\Omega_{\chi} h^2 < 0.05$). The dashed line delimits the upper
frontier of the scatter plot, when the inputs of Ref. \cite{efo}
are used. The solid contour denotes the 3$\sigma$ annual--modulation
region of Ref. \cite{damalast} (with the specifications given in the text). 
}
\end{figure}

\newpage
\begin{figure}[t]
\hbox{
\psfig{figure=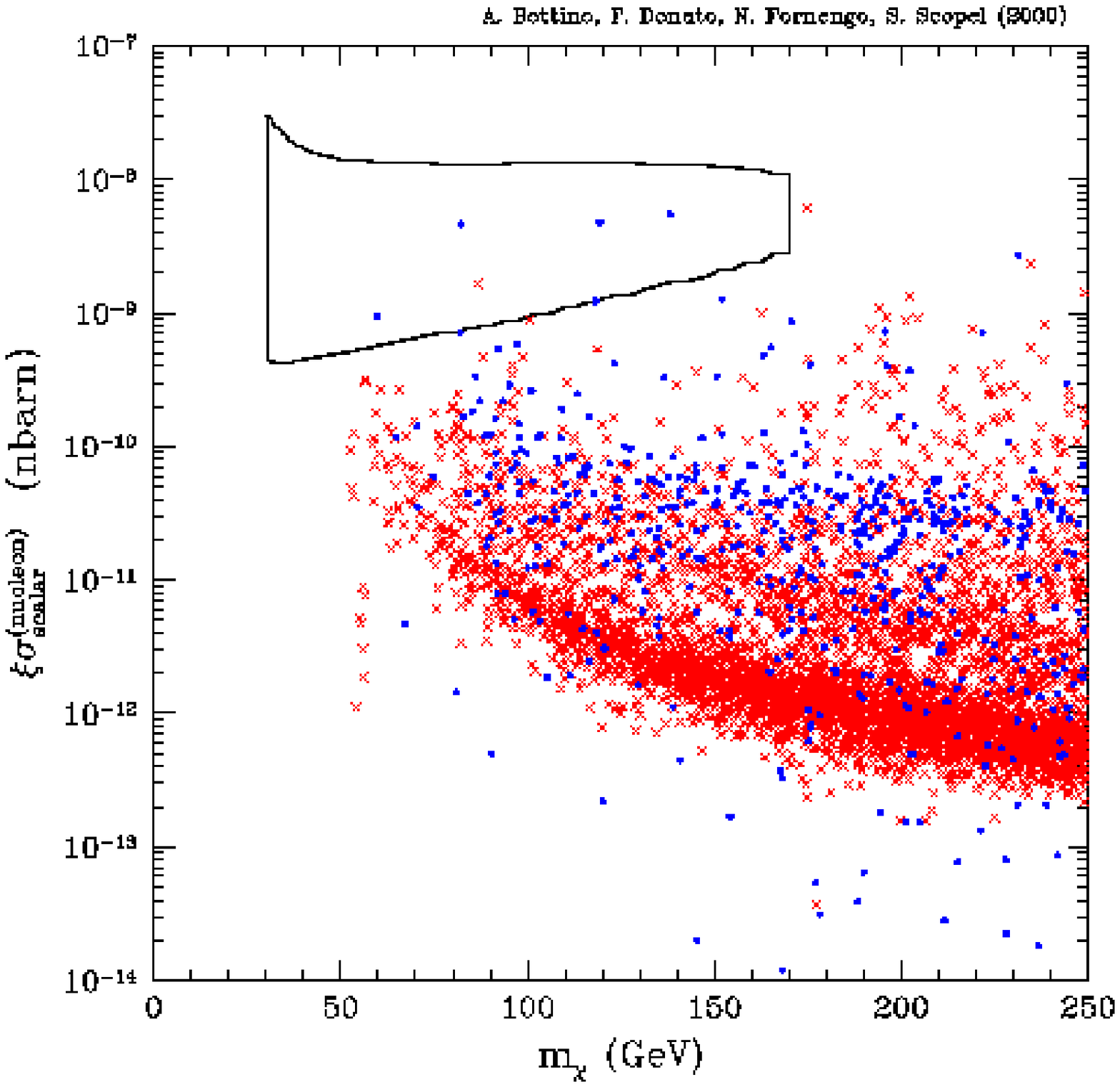,width=8.2in,bbllx=40bp,bblly=180bp,bburx=700bp,bbury=660bp,clip=}
}
{
FIG. 5b.
Same as in Fig. 5a in case of nuSUGRA.
}
\end{figure}

\newpage
\begin{figure}[t]
\hbox{
\psfig{figure=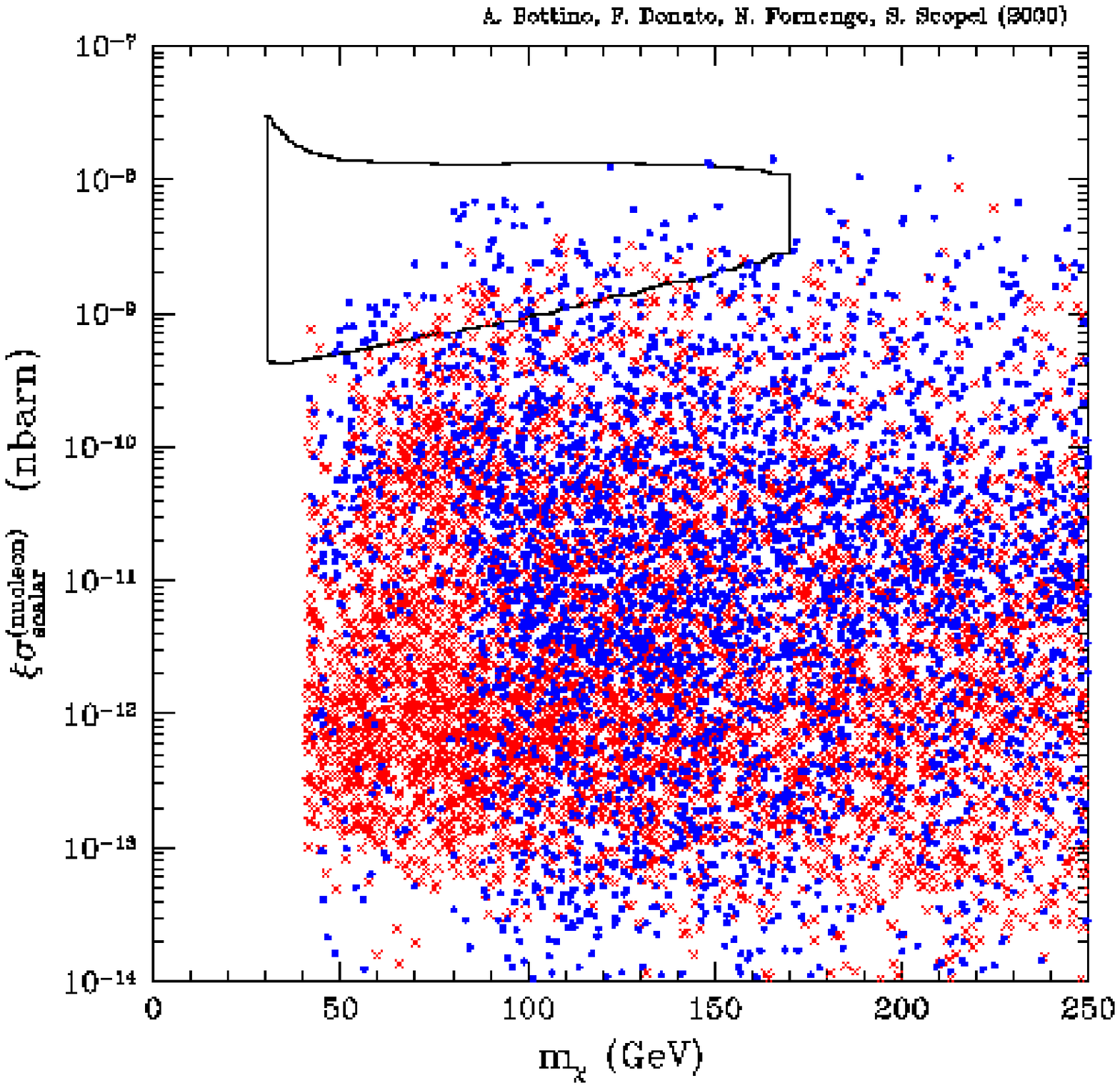,width=8.2in,bbllx=40bp,bblly=180bp,bburx=700bp,bbury=660bp,clip=}
}
{
FIG. 5c.
Same as in Fig. 5a in case of effMSSM.
}
\end{figure}

\end{document}